ARTICLE  OPEN

# Strong two-dimensional plasmon in Li-intercalated hexagonal boron-nitride film with low damping

Ivor Lončarić[1], Zoran Rukelj[2], Vyacheslav M. Silkin[3,4,5] and Vito Despoja[3,4,6]

The field of plasmonics seeks to find materials with an intensive plasmon (large plasmon pole weight) with low Landau, phonon, and other losses (small decay width). In this paper, we propose a new class of materials that show exceptionally good plasmonic properties. These materials consist of van der Waals stacked "plasmon active" layers (atomically thin metallic layers) and "supporting" layers (atomically thin wide band gap insulating layers). One such material that can be experimentally realized—lithium intercalated hexagonal boron-nitride is studied in detail. We show that its 2D plasmon intensity is superior to the intensity of well-studied Dirac plasmon in heavy doped graphene, which is hard to achieve. We also propose a method for computationally very cheap, but accurate analysis of plasmon spectra in such materials, based on one band tight-binding approach and effective background dielectric function.

## INTRODUCTION

Quasi-two-dimensional crystals consisting of one isolated atomic layer, such as graphene or hexagonal boron nitride (hBN),[1–4] or consisting of few (covalently bound) atomic layers, such as transition metal dichalcogenides,[5,6] are widely investigated as they possess multitude of useful properties. While some of them have tunable concentration of charge carriers that can be exploited in plasmonics, photonics, and transformation optics,[7–10] others have direct bandgap which makes them suitable for electronics, optoelectronics, light emitters, detectors, and photovoltaic devices.[8,11–20]

Alkaline or alkaline earth intercalated graphene has also recently been extensively studied theoretically and experimentally.[21–31] This kind of graphene intercalates supports two 2D plasmons; Dirac and acoustic plasmons, which are a consequence of hybridization between the two 2D plasmas lying in graphene and metallic planes.[32] The metallic layer provides strong natural doping of graphene layers resulting in very strong Dirac plasmon. However, intraband and interband electron–hole excitations within metallic or graphene planes cause direct (Landau) or indirect (e.g., via phonons or impurities[31,33]) Dirac plasmon relaxation.

Other graphene-based materials, such as the graphene nanoribbons,[34,35] the twisted graphene bilayer,[36] or the graphene on SiC or SiO₂ substrates[37,38] also supports a variety of 2D plasmons or (hybridized) plasmon/phonon modes.[39] However, their experimental observation is very difficult,[40] either because of the very small doping or because of the experimental difficulties related to the low structural stability of such systems. The question which here arises is whether there exists the related 2D materials which would support strong isolated 2D plasmon?

By careful stacking of atomically thin crystal layers one could engineer the heterostructures or van der Waals layered crystals (vdWcrys) with desired properties.[41,42] Here, we propose that the vdWcrys ideal for plasmonic applications can be obtained by successive stacking of atomically thin metallic layers that we call "plasmon active" layers and atomically thin wide band gap insulating layers that we call "supporting" layers. In this way, low plasmon losses are obtained as the metallic band in "plasmon active" layer (which provides 2D plasmon) is placed in a wide band gap. This reduces both 2D plasmon relaxation to direct interband electron–hole excitations (Landau damping), and also relaxation to indirect interband electron–hole transitions, caused by electron–phonon, electron–impurity, or electron–electron scattering.[43] However, Drude relaxation of 2D plasmon to intraband transitions still remains. The dominant intraband relaxation channel comes from interaction with LO phonons in supporting layers which are here at vdW separations (≈3 Å) from metallic monolayer (ML) and probably weakly interact with 2D plasmon. Therefore, the only 2D plasmon decay channel comes from interaction with longitudinal acoustic (LA) phonons in the metallic plane, which is also estimated to be small. For example, using the jellium model predictions, the electron–LA phonons coupling constant in the lithium metallic plane[44] is one order of magnitude weaker than the analog coupling constant in graphene calculated from the first principles.[45] This suggests generally low phonon losses in this kind of vdWcrys.

The material we study here in detail is lithium intercalated hBN (LiB$_2$N$_2$). It consists of two parallel hexagonal boron-nitride (hBN) layers ("supporting" layers) encapsulating one lithium layer ("plasmon active" layer) where Li atoms occupy hBN hollow sites in order 2 × 2.

[1]Ruđer Bošković Institute, Bijenička 54, HR-10000 Zagreb, Croatia; [2]Department of Physics, Faculty of Science, University of Zagreb, Bijenička 32, HR-10000 Zagreb, Croatia; [3]Donostia International Physics Center (DIPC), P. Manuel de Lardizabal, 4, 20018 San Sebastián, Spain; [4]Departamento de Fisica de Materiales, Facultad de Ciencias Químicas, Universidad del Pais Vasco UPV/EHU, Apto. 1072, 20080 San Sebastián, Spain; [5]IKERBASQUE, Basque Foundation for Science, 48011 Bilbao, Spain and [6]Institute of Physics, Bijenička 46, HR-10000 Zagreb, Croatia
Correspondence: Vito Despoja (vito@phy.hr)

Published in partnership with FCT NOVA with the support of E-MRS

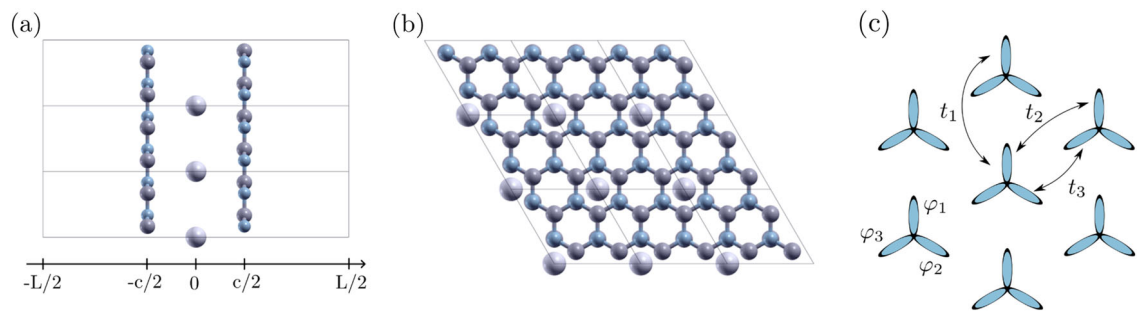

**Fig. 1** The schematic representation of $LiB_2N_2$ crystal structure. **a** The lithium ("plasmon active") layer occupies $z = 0$ plane and two parallel hBN ("supporting") layers occupy $z = \pm c/2$ planes. **b** Lithium atoms occupy hBN hollow sites in the order $2 \times 2$. **c** TBA hoping parameters

The concept of strong 2D plasmon proposed here does not dependent neither on alkali metal nor on particular superstructure or number of hBN layers. Although, here we focus on Li $2 \times 2$ superstructure any other combination would give qualitatively the same conclusions, even hBN could be replaced by other 2D insulator/semiconductor. Li-intercalated hBN was synthesized and experimentally investigated in refs. [46,47] K intercalated hBN[48] shows $2 \times 2$ monolayer superstructure (as in graphene case too). In ref. [49], the authors studied hBN/Au/Ni(111) doped by K and Li atoms. They show that K prefers to intercalate in between hBN and Au/Ni(111) and Li to adsorb on top of hBN (the same happens for graphene/Au/Ni(111)). Li and Cs can also be intercalated in hBN/Ir(111) system.[50] Metallic surface destroys Li σ band, which supports 2D plasmon, and, therefore, Li would have to be encapsulated with another hBN layer. However, these systems are relevant for device integration.

hBN layers ensure wide band gap (4.5 eV in this calculation) and lithium plane ensures metallic (parabolic) σ band placed in band gap resulting in the formation of isolated (almost freestanding) two-dimensional plasma. The plasmonics in artificial freestanding lithium layer was theoretically studied in ref. [51] showing intensive 2D plasmon. However, this phenomenon of freestanding "metallic" 2D plasmon has not been experimentally detected yet. Therefore, this theoretical investigation encourages such measurements, taking into account that Li-intercalated hBN is known to exist.[46,47]

In this paper, we calculate plasmonic properties of $LiB_2N_2$ ML and compare them to properties of self-standing lithium ML and heavy doped graphene. The "tunable" Dirac plasmon in graphene has been intensively studied due to its many possible applications, however, for experimentally feasible dopings it is usually very weak and broad resonance.[52] Although doping increases the Dirac plasmon intensity, the heavy dopings, for which it becomes intensive and useful mode, are difficult or even impossible to achieve.[35,38]

We show that in proposed realistic material—lithium intercalated hBN, 2D plasmon is not as intensive as in self-standing Li-ML due to screening from hBN, but it is still superior to Dirac plasmon in heavy doped graphene that is hard to produce. We also show that correct plasmon energy and intensity can be only obtained if crystal local field effects in the perpendicular direction are included in the calculation. Crystal local field effects in this system can be efficiently treated if the dielectric response of the lithium layer is described by the 2D response function and dielectric response of two hBN layers by an effective background dielectric function that includes effects coming from spatially dependent Coulomb interaction between separated layers and perpendicular local field effects in independent hBN layers.

We also propose the fast and accurate analysis of energy loss function (ELF) in related layered systems where the dielectric response of plasmon active layer can be modeled by one band tight binding approximation (TBA), here Li-$\sigma_1$ parabolic band.

Response function of "supporting" layers can be modeled by static effective background dielectric function, which for wide band gap insulating layers has a simple analytical parametrization.

## RESULTS AND DISCUSSION

The $LiB_2N_2$-ML crystal consists of lithium ("plasmon active") layer placed in $z = 0$ plane and two parallel hBN layers ("supporting") layers placed in $z = \pm c/2$ planes, as shown in Fig. 1a. Lithium atoms occupy hBN hollow sites in the order $2 \times 2$, as shown in Fig. 1b.

Before analyzing, for plasmonics relevant, low energy ELF in $LiB_2N_2$-ML, we study plasmonic properties in the self-standing Li-ML. The Li-ML crystal structure is the same as in $LiB_2N_2$-ML but with removed hBN monolayers. Figure 2a shows the Li-ML band structure along the high symmetry directions $\Gamma \to K \to M \to \Gamma$. The red patterns, denoted as $\sigma_1$, $\sigma_2$, and $\sigma_3$, show the contribution of lithium $2[s, p_x,$ and $p_y]$ orbitals (Li-σ bands) to the band structure. Two lithium π bands are also denoted by white arrows. The blue dots show the σ-TBA model band structure (see Fig. 1c). The TBA bands lie on top of ab initio Li-σ bands, except around M point. However, energies there are already too high for plasmonics in Terahertz (THz) and infra-red (IR) frequency range ($\omega < 1.5$ eV). The most relevant is the lowest $\sigma_1$ band that crosses the Fermi level for which ab initio and TBA bands agree well. Moreover, the TBA calculation confirms that the $\sigma_1$ band is entirely parabolic $\hbar^2 K^2 / 2m^*_{\sigma_1}$ with the effective mass $m^*_{\sigma_1} = 1.15$.

Figure 2b shows ELF, Eq. (6), in Li-ML obtained from dielectric function (7), where $\varepsilon_B = 1$ and Eq. (8) represents the independent electrons response function in Li-ML. The spectra show very intensive 2D plasmon which mostly originates from intraband electron–hole transitions within $\sigma_1$ band. The white dotted line denotes the upper edge of intraband electron–hole pair continuum in parabolic band approximation ($\omega_+ = (Q^2 + 2k_F Q)/2m^*_{\sigma_1}$). The energy of $\sigma_1$ band in Γ point is $E_{\sigma_1 \Gamma} = -0.96$ eV and thus, the Fermi wave vector is $k_F = 0.285$ a.u. It can be clearly seen how weak yellow intensity pattern, corresponding to intraband continuum, fill the region exactly below $\omega_+$.

In order to demonstrate the limits of validity of simple one band model, especially in important THz and IR frequency range, the green dotted line in Fig. 2b shows the plasmon dispersion relations obtained by solving Eq. (11) where only intraband transitions within Li-ML $\sigma_1$ band are included in the response function (8). This dispersion relation agrees well with the 2D plasmon intensity pattern until $Q \approx 0.1$ a.u. For larger Q's, it starts to deviate such that it decreases and finally shows negative dispersion. This deviation is because the one band model wrongly considers energetically the lowest band, instead of symmetrically equivalent band (which in $K \to M$ region increases, not decreases). This disables higher energy intraband excitations and therefore lowers the 2D plasmon energy. The blue dotted line shows the 2D

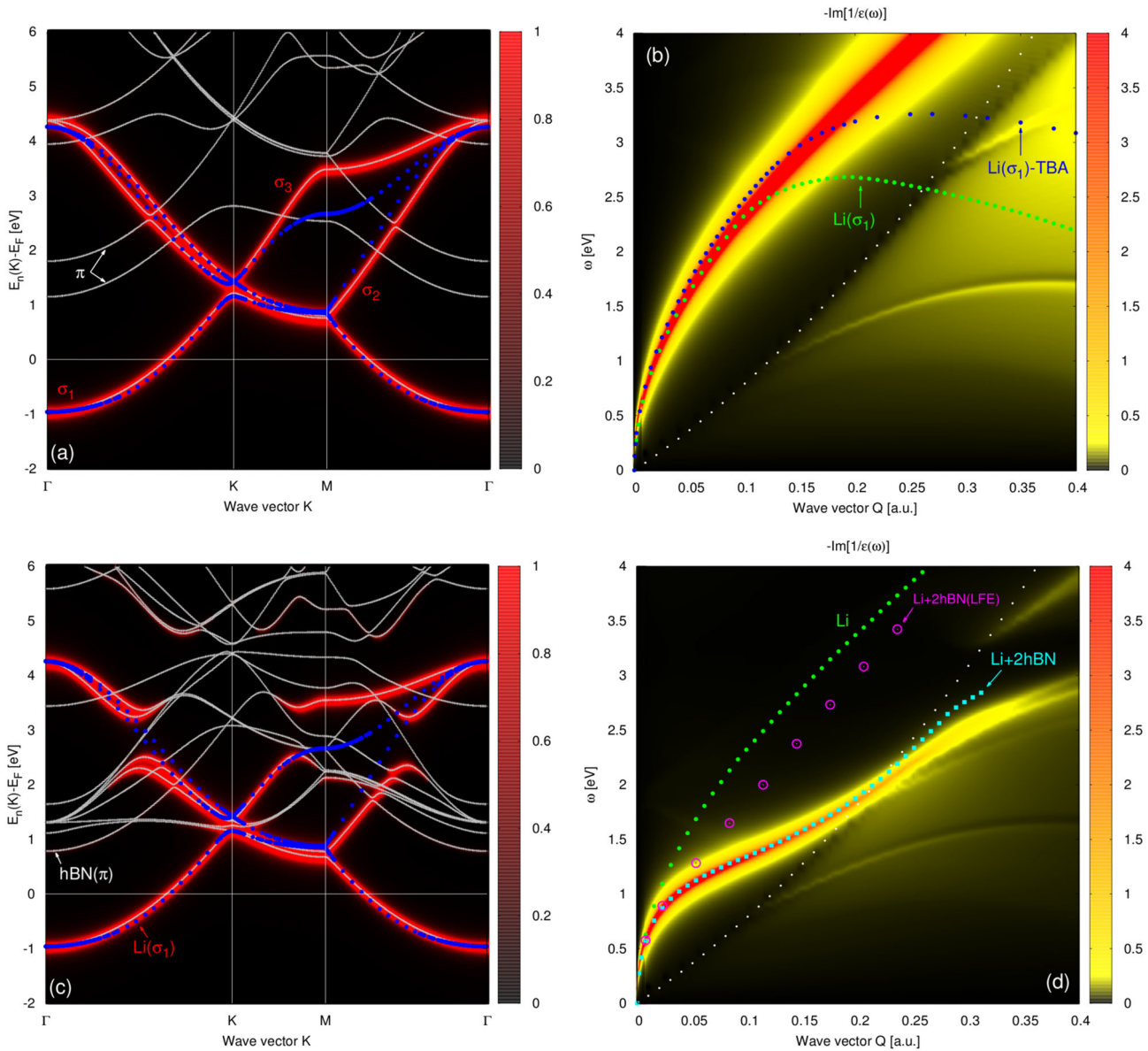

Fig. 2  **a** Li-ML band structure. The red patterns show the intensity of lithium σ bands. The blue dots show band structure obtained using σ-TBA model. **b** The ELF in Li-ML. The white dots show the upper edge $\omega_+ = (Q^2 \pm 2k_F Q)/2m^*_{\sigma_1}$ of intraband electron–hole continuum in parabolic ($E_K = K^2/2m^*_{\sigma_1}$) $\sigma_1$ band, where $m^*_{\sigma_1} = 1.15$ and corresponding Fermi wave vector is $k_F = 0.285$ a.u. The green dotted line shows the 2D plasmon dispersion relation for which only transitions within Li-ML $\sigma_1$ band are included. The blue dotted line shows the same 2D plasmon dispersion relation obtained within $\sigma_1$-TBA model. **c** The LiB$_2$N$_2$-ML band structure. Blue dots show the same as in **a**. **d** The ELF in LiB$_2$N$_2$-ML. The dispersion relations (11) in Li-ML (green dots), in Li + 2hBN model system (cyan squares) and in Li + 2hBN(LFE) model system (magenta circles). White dots show the same as in **b**

plasmon dispersion relation obtained using $\sigma_1$-TBA model. As can be seen in the figure, for $Q < 0.05$ a.u. (which would correspond to THz and IR frequency range, $\omega \leq 1.5$ eV, considering the 2D plasmon energy) the $\sigma_1$-TBA model agrees well with the ab initio result. This suggests that simple $\sigma_1$-TBA model is capable to reproduce THz and IR plasmonics in Li-ML.

Figure 2c shows the full LiB$_2$N$_2$-ML band structure where the red intensity patterns show the intensity of Li-σ bands. In comparison with Li-ML band structure, it is obvious that the presence of two hBN monolayers distorts the lithium $\sigma_2$ and $\sigma_3$ bands, however, the $\sigma_1$ band remains unaffected. Also a multitude of new bands appear which are mostly build from hBN π orbitals. The blue dotted lines show the σ-TBA band structure for comparison.

Figure 2d shows the energy loss spectrum (6) in LiB$_2$N$_2$-ML obtained from dielectric function (7), where $\varepsilon_B = 1$ and Eq. (8) represents independent electron response function of LiB$_2$N$_2$-ML. For comparison, the green dotted line represents 2D plasmon dispersion relation $\omega_{pl}(\mathbf{Q})$ in Li-ML. Furthermore, black and red solid lines in Fig. 3 show the plasmon pole weights (12) as a function of wave vector **Q** in Li-ML and LiB$_2$N$_2$-ML, respectively. Clearly, the two hBN adlayers strongly reduce 2D plasmon energy. Unfortunately, the intensity of the 2D plasmon is also suppressed. A similar effect (Dirac plasmon suppression) occurs in graphene, see ref. [53]. Note, additionally, that the intensity of $\sigma_1$ intraband electron–hole continuum is reduced.

In order to understand this dramatic modification of Li-ML 2D plasmon, we use the following simplified model. In the self-standing Li-ML, the independent electrons charge density fluctuations, described by the response function $\chi^0$, interact via bare Coulomb interaction $v_Q$ and the total screened Coulomb interaction can be written as $W = v_Q/(1 - \chi^0 v_Q)$. Let us suppose now that we add two hBN layers placed in $z_1 = -c/2$ and $z_2 = c/2$ planes (see Fig. 1a). This model system that we will call Li+2hBN, consists of the Li-ML and two hBN layers which are at the DFT level of calculation (considering large vdW separation) treated as fully independent (nonoverlapping) layers. These layers can interact only by the long-range Coulomb interaction (a similar approach was used in ref. [54]). If we describe two hBN layers by the static 2D dielectric function $\varepsilon_{2hBN}(Q)$ then the bare Coulomb interaction propagator in Li layer becomes screened, i.e., it becomes

$$v_Q \to v_Q/\varepsilon_{2hBN}(Q) \quad (1)$$

and the total dynamically screened Coulomb interaction in Li layer becomes

$$W = v_Q/(\varepsilon_{2hBN} - \chi^0 v_Q). \quad (2)$$

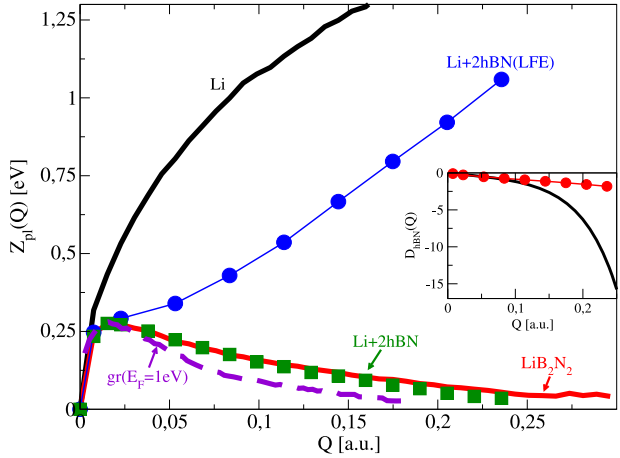

**Fig. 3** The plasmon pole weights (12) in Li-ML (black solid), $LiB_2N_2$-ML (red solid), Li+2hBN (green squares), Li+2hBN(LFE) (blue dots), and in doped ($E_F = 1$ eV) graphene (violet dashed). The inset shows the static hBN surface excitation propagator calculated without LFE (black solid) and with LFE (red dots)

Due to the fact that the hBN layer has a large band-gap (about 4.5 eV in this calculation) we estimate that the static approximation is valid for $\omega < 3.5$ eV, which covers the energy interval relevant for this investigation. Moreover, linear approximation $\varepsilon_{2hBN}(Q) = 1 + \alpha_{2hBN}Q$ is valid in almost whole shown wave vector interval. The static polarizability $\alpha_{2hBN}$ of the two hBN layers can be obtained from single hBN layer static polarizability $\alpha_{hBN}$, simply as $\alpha_{2hBN} = \alpha_{hBN}$. All in all, in this simplified model the introduction of two 2D hBN layers changes the effective background 2D dielectric function $\varepsilon_B$ that enters Eq. (7) as

$$\varepsilon_B = \varepsilon_{2hBN}(Q) = 1 + 2\alpha_{hBN}Q. \quad (3)$$

From separate ab initio calculation for isolated hBN layer we obtain $\alpha_{hBN} = 12.6$ a.u.

The cyan squares in Fig. 2d show the plasmon dispersion relation (11) in the Li+2hBN model system obtained from Eq. (7), Li-ML response function (8) and the effective background 2D dielectric function (3). It can be seen that plasmon dispersion relation from the simplified model excellently agrees with $LiB_2N_2$-ML ELF intensity pattern up to $Q \approx 0.25$ a.u. Moreover, the green squares in Fig. 3 that show the plasmon pole weight in Li+2hBN model system, also excellently agree with plasmon pole weight in the entire $LiB_2N_2$-ML system (red line). This success of the proposed model system proves that electronic densities of the three systems (Li-ML and two hBN layers) weakly overlap and that they interact only via long-range Coulomb interaction, which actually confirms the vdW character of the crystal plane stacking. More importantly, the success of the model system proves that the strong reduction of the Li-ML 2D plasmon energy and intensity is entirely due to the static screening from surrounding hBN layers.

Since the thickness of the complete system ($LiB_2N_2$-ML) is more than $c = 6.3$ Å, the fully 2D approximation of its response that was used up to now is valid only for $Q < \frac{1}{c} \approx 0.15$ Å$^{-1} \approx 0.08$ a.u. Therefore, for wave vectors $Q \geq 0.05$ a.u. one should include the crystal local field effects or dispersivity of the $LiB_2N_2$-ML nonlocal response in perpendicular (z) direction. Since the inclusion of the crystal local field effects for the entire $LiB_2N_2$ system is computationally very demanding and due to the fact that three layers weakly overlap, in the following we provide computationally very cheap local field effects calculation based on the above discussed Li+2hBN model system.

A new model system, called Li+2hBN(LFE), consists of Li-ML placed in $z = 0$ plane that is still treated as intrinsically 2D system and two hBN layers placed in $z = -c/2$ and $z = c/2$ planes with response functions that become nonlocal function of $z$ and $z'$. In other words, each hBN layer is described by nonlocal independent electrons response function $\chi^0(\mathbf{Q}, \omega, z, z')$. In the Fourier expansion of hBN response function $\chi^0(\mathbf{Q}, \omega, z, z') = \frac{1}{L}\sum_{G_z,G_z'} \chi^0_{G_z,G_z'}(\mathbf{Q}, \omega)e^{iG_z z - iG_z' z'}$, where $G_z = 2\pi n/L$; $n = ..., -2, -1, 0, 1, 2, ...$ represent the reciprocal lattice vectors in $z$ direction, we have used the energy cut-off of 5 Hartrees which corresponds to the matrix of dimension $23 \times 23$. Introduction of such hBN layers causes that the bare Coulomb propagator $v_Q$ in Li layer is screened as[55]

$$v_Q \to v_Q \frac{1 + D_{hBN}(Q)e^{-Qc}}{1 - D_{hBN}(Q)e^{-Qc}} = v_Q/\varepsilon^{LFE}_{2hBN}, \quad (4)$$

where $D_{hBN}(Q)$ represents the static hBN surface excitation propagator. Calculation of this propagator in terms of nonlocal response function $\chi^0(\mathbf{Q}, \omega, z, z')$ was described in refs. [32,56–58]. With replacement (4), the total dynamically screened Coulomb interaction in Li layer becomes $W = v_Q/(\varepsilon^{LFE}_{2hBN} - \chi^0 v_Q)$, i.e., the effective background 2D dielectric function becomes

$$\varepsilon_B = \varepsilon^{LFE}_{2hBN}(Q) = \frac{1 - D_{hBN}(Q)e^{-Qc}}{1 + D_{hBN}(Q)e^{-Qc}}. \quad (5)$$

As already mentioned, large hBN band gap ensures that the static approximation of propagator $D$ is valid in the frequency interval of interest here. Great advantage of such modeled perpendicular local field effects is that they now give proper Coulomb interaction between spatially separated (for $c/2$) Li and hBN layers and in the same time they ensure perpendicular dispersivity of hBN nonlocal response.

Magenta circles in Fig. 2d show the dispersion relation and blue dots in Fig. 3 show the plasmon pole weight of 2D plasmon in Li+2hBN(LFE) system. It is obvious that the inclusion of local field effects recuperates the plasmon energy and intensity. This is reasonable considering that the inclusion of local field effects (4) reduces screening coming from hBN layers, for two reasons. First, the inclusion of local field effects implies that hBN layers are at large spatial separations $c/2 \approx 3.2$ Å from Li atomic layer, which reduces screening coming from hBN layers for larger wave vectors (e.g., $Q > 2/c \approx 0.1$ a.u.). Second, spatial dispersivity of hBN nonlocal response strongly weakens its screening for larger wave vectors ($Q > 0.1$ a.u.), as can be seen in the inset of Fig. 3 which shows hBN propagator $D$ without local field effects (black solid line) and with local field effects (red dots) included.

Violet dashed line in Fig. 3 shows the Dirac plasmon pole weight in heavy doped graphene ($E_F = 1$ eV), for comparison. The ELF in graphene is calculated using the same procedure (6), (7), and $\varepsilon_b = 1$ where 2D response function (8) is calculated as described in ref. [55]. For $Q < 0.02$ a.u. which, considering the 2D plasmon energy in Li+2hBN(LFE) system, corresponds to THz to IR frequency range, weight of Dirac graphene plasmon and 2D plasmon in Li+2hBN(LFE) system coincide. However, for larger wave vectors 2D plasmon weight increases and becomes much higher than the Dirac plasmon weight which decreases. For example, for $Q \approx 0.05$–$0.075$ a.u., which corresponds to IR frequency range, 2D plasmon is 2–4 times more intensive than Dirac plasmon, and for $Q \approx 0.075$–$0.15$ a.u. (visible frequency range), 2D plasmon is 4–15 times more intensive than Dirac plasmon. This suggests that 2D plasmon in Li+2hBN(LFE) system is superior to Dirac plasmon in heavy doped graphene in IR and visible frequency range.

The analysis performed here also anticipate the possibility of fast and accurate calculation of ELF in related layered systems. Namely, the plasmon active layer can be modeled simply by TBA response function (8) and "supporting" layers by some effective background dielectric function $\varepsilon_b$. For example, for the wave vectors $Q < 0.15$ a.u., the ELF of entire $LiB_2N_2$-ML can be modeled simply by $\sigma_1$-TBA response function (8) and for the effective background dielectric function one can take Eq. (5). As can be seen from the inset of Fig. 3, for hBN surface excitations propagator one can use linear approximation $D_{hBN}(Q) \approx D(Q = 0) + \beta_{hBN} Q$, where $D(Q = 0) = 0$ and $\beta_{hBN} = -8.4$ a.u.

In this paper, we have proposed the new class of plasmonic materials that consist of vdW stacked "plasmon active" layers (atomically thin metallic layer) and "supporting" layers (2D wide band gap insulating layers). The particular system we have investigated is $LiB_2N_2$-ML where lithium layer represents plasmon active layer and two hBN layers are insulating supporters. 2D plasmon energy and intensity of this system are superior to well known Dirac plasmon in heavy doped graphene. Moreover, experimental realization of proposed system should be rather simple in comparison with heavy doped graphene.

We have shown that crystal local field effects have to be included in the modeling to obtain correct 2D plasmon energy and intensity. We have also proposed computationally cheap and accurate methodology for the inclusion of crystal local field effects. In this methodology, the response of plasmon active layer is approximated by 2D response function. The response of supporting layers is approximated by effective background dielectric function that implements spatially dependent Coulomb interaction between separated layers and perpendicular crystal local field effects in supporting layers. Fast and accurate calculation of ELF in long-wavelength limit can be performed if the plasmon active layer is simply modeled by the one band TBA response function and the effective background dielectric function is expressed in terms of static linear surface excitations propagator $D(Q) = D(0) + \beta Q$. This allows fast semi-analytical modeling of ELF by using just few parameters.

## METHODS

### Ab-initio approach

For the calculation of the Kohn–Sham wave functions $\phi_{n\mathbf{K}}$ and energy levels $E_{n\mathbf{K}}$, i.e., the band structure, we use the plane-wave density functional theory (DFT) code Quantum ESPRESSO.[59] The core–electron interaction is approximated by the norm-conserving pseudopotentials,[60] and the exchange correlation (XC) potential by the vdW-DF-cx functional.[61,62] For the $LiB_2N_2$-ML unit cell constant we use $a_{uc} = 5.02$ Å and superlattice constant in $z$ direction is $L = 14$ Å. The ground state electronic densities of the $LiB_2N_2$-ML is calculated by using the $10 \times 10 \times 1$ Monkhorst–Pack $\mathbf{K}$-point mesh[63] of the first Brillouin zone (BZ). For the plane-wave cut-off energy we choose 820 eV. Structural optimization was performed until a maximum force below 0.002 eV/Å was obtained. For the equilibrium separation between hBN layers encapsulating Li we obtain $c = 6.3$ Å.

The probability density $P(\mathbf{Q},\omega)$ for the parallel momentum transfer $\mathbf{Q}$ and the energy loss $\omega$ of the reflected electron in Electron energy loss spectroscopy (EELS) experiments[64] is proportional to the imaginary part of the dynamically screened Coulomb potential $W = v_Q/\varepsilon$, i.e.,

$$P(\mathbf{Q},\omega) \propto -\mathbf{1}/\varepsilon(\mathbf{Q},\omega), \quad (6)$$

which is called the ELF. If vdWcrys is considered as a fully 2D system its total dielectric function can be written as

$$\varepsilon(\mathbf{Q},\omega) = \varepsilon_B(\mathbf{Q},\omega) - \mathbf{v_Q}\chi^0(\mathbf{Q},\omega), \quad (7)$$

where $\varepsilon_B$ represents effective background 2D dielectric function and the 2D independent electrons response function is given by

$$\chi^0(\mathbf{Q},\omega) = \frac{2}{S} \sum_{\mathbf{K} \in \mathbf{S.B.Z.}} \sum_{\mathbf{n,m}} \frac{f_{n\mathbf{K}} - f_{m\mathbf{K}+\mathbf{Q}}}{\hbar\omega + i\eta + E_{n\mathbf{K}} - E_{m\mathbf{K}+\mathbf{Q}}} \rho_{n\mathbf{K},m\mathbf{K}+\mathbf{Q}} \rho^*_{n\mathbf{K},m\mathbf{K}+\mathbf{Q}}. \quad (8)$$

Here $f_{n\mathbf{K}} = \left[e^{(E_{n\mathbf{K}} - E_F)/kT} + 1\right]^{-1}$ is the Fermi–Dirac distribution at temperature $T$ and the charge vertices are

$$\rho_{n\mathbf{K},m\mathbf{K}+\mathbf{Q}} = \langle \phi_{n\mathbf{K}} | e^{-i\mathbf{Q}\mathbf{r}} | \phi_{m\mathbf{K}+\mathbf{Q}} \rangle_\Omega \quad (9)$$

where $\mathbf{Q}$ is the momentum transfer vector parallel to the $x$–$y$ plane and $\mathbf{r} = (\boldsymbol{\rho}, z)$ is a 3D position vector. Integration in (9) is performed over the normalization volume $\Omega = S \times L$, where $S$ is normalization surface. Plane wave expansion of the wave function has the form $\phi_{n\mathbf{K}}(\boldsymbol{\rho}, z) = \frac{1}{\sqrt{\Omega}} e^{i\mathbf{K}\boldsymbol{\rho}} \sum_{\mathbf{G}} C_{n\mathbf{K}}(\mathbf{G}) e^{i\mathbf{G}\mathbf{r}}$, where $\mathbf{G} = (\mathbf{G}_\parallel, G_z)$ are 3D reciprocal lattice vectors and the coefficients $C_{n\mathbf{K}}$ are obtained by solving the Kohn–Sham equations self-consistently.

The independent electron response function (8) is calculated by using $101 \times 101 \times 1$ $\mathbf{K}$-point mesh sampling which corresponds to 10,303 Monkhorst–Pack special $\mathbf{k}$-points in the Brillouin zone. The damping parameter we use is $\eta = 30$ meV and temperature is $kT = 10$ meV. The band summation is performed over 100 bands, which proved to be sufficient for proper description of the electronic excitations up to 20 eV. As for the $\mathbf{K}$-point mesh used in this calculation the minimum transfer wave vector $|\mathbf{Q}|$ is $Q_c = 0.0076$ a.u.$^{-1}$. For $|\mathbf{Q}| < Q_c$ the independent electrons response functions is calculated using optical conductivity $\sigma(\omega) = \sigma^{intra}(\omega) + \sigma^{inter}(\omega)$, obtained in the strict $\mathbf{Q} = \mathbf{0}$ limit, as[65]

$$\chi^0(\mathbf{Q} < Q_c, \omega) = L \frac{\mathbf{Q}^2}{i\omega} \sigma(\omega). \quad (10)$$

To obtain this expression we combined the continuity equation, $\omega\rho = \mathbf{Q} \cdot \mathbf{j}$, the Ohm's law for the induced current, $\mathbf{j} = \sigma(\omega)\mathbf{E}^{tot}$, the total electric field expressed as $\mathbf{E}^{tot} = -i\mathbf{Q}\phi^{tot}$, and the relation defining the response function, $\rho(Q,\omega) = \chi^0(Q,\omega)\phi^{tot}(Q,\omega)$. The Drude or intraband $\sigma^{intra}$ and interband $\sigma^{inter}$ conductivities are explicitly given by Eqs. (7)–(12) of ref.[55]. For calculation of $\sigma^{intra}$ and $\sigma^{inter}$ we used the same parameters as for calculation of independent electrons response function (8) and $\eta_{intra} = \eta_{inter} = 30$ meV. Finally the ELF in optical limit ($|\mathbf{Q}| < Q_c$) is obtained using (6), (7), and (10).

For the proper quantitative analysis of plasmon, it is crucial to accurately define its energy (plasmon dispersion relation) and strength (plasmon pole weight). The 2D plasmon dispersion relation $\omega_{pl}(\mathbf{Q})$ is defined as poles of screened Coulomb interaction $W = v_Q/\varepsilon$, or zeros of dielectric function (7), i.e.,

$$\varepsilon[\mathbf{Q}, \omega_{pl}(\mathbf{Q})] = 0. \quad (11)$$

The plasmon pole weight is defined as the residue of screened Coulomb interaction $W$ around the plasmon pole $\omega_{pl}(\mathbf{Q})$, i.e.,

$$Z_{pl}(Q) = \left[\frac{\partial \varepsilon(\mathbf{Q},\omega)}{\partial \omega}\right]^{-1} \bigg|_{\omega_{pl}(\mathbf{Q})}. \quad (12)$$

It should be noted here that plasmon pole weight (12) depends on both Landau damping and screening mechanisms in the system. In electron–hole forbidden region (no Landau damping) plasmon spectral weight is defined only by screening strength and the plasmon width is equally broad and set by the damping parameter $\eta$.

For separate hBN layer ground state calculation we used the norm-conserving pseudopotentials,[60] and the LDA XC potential. For the hBN layer unit cell constant we use value of $a_{uc} = 4.746$ a.u. and the neighboring hBN layers (in supercell) are separated by $L = 5a_{uc} = 23.73$ a.u. The ground state electronic density is calculated by using $12 \times 12 \times 1$ Monkhorst–Pack $\mathbf{K}$-point mesh sampling. For the plane-wave cut-off energy we choose 680 eV. The hBN-ML independent electron response function $\chi^0$ (8) is calculated by using $201 \times 201 \times 1$ Monkhorst–Pack $\mathbf{K}$- point mesh sampling. The damping parameter we use is $\eta = 30$ meV and the band summation is performed over 30 bands.

### TBA approach

Since we show that at the DFT level of calculation the problem can be separated to 2D electron gas in Li plane plus simple parametrized screening coming from two hBN layers, we also consider tight binding parametrization of 2D electron gas. This approach would enable computationally cheap screening of plasmonic properties in related materials. In this way, one could also include effects such as impurity and phonon-assisted 2D plasmon damping and perform more sophisticated many-body calculations beyond RPA.

Here we describe $sp^2$ hybrid tight binding approach ($\sigma$-TBA) used to describe valence $\sigma$ bands in self-standing lithium monolayer Li-ML. Due to the hexagonal symmetry of the underlying lattice it is natural to construct the $sp^2$ hybrids on each lithium atom using $2s$, $2p_x$, and $2p_y$ atomic orbitals. These hybrids $\varphi_i$ ($i = 1,2,3$) are constructed from the atomic orbitals by the unitary Pauling transformation, and are depicted schematically in Fig. 1c. In the case of the trigonal symmetry, the crystal field effects are such that all three hybrids $\varphi_i$ have the same energy

$$E_\varphi^0 = \frac{1}{3}E_s^0 + \frac{2}{3}E_p^0 \qquad (13)$$

with the $E_s^0$ and $E_p^0$ being the atomic energies of the $2s$ and $2p$ orbitals, respectively. The total crystal Hamiltonian contains the atomic Hamiltonian and the effective single particle interaction called the crystal potential. This potential is responsible for the change of the atomic energies of the hybrids compared to the free atomic ones, and for the tunneling of the electron between the hybrids located on the neighboring lattice positions $\rho_i$. Here, we consider electron hopping only to the nearest neighbors and neglect any overlap of the neighboring hybrids. In this case the TBA Hamiltonian is

$$\hat{H} = \sum_{i,j \in \varphi} \sum_{\boldsymbol{\rho}, \mathbf{R}, \sigma} H^{ij}(\boldsymbol{\rho}) c^+_{i, \mathbf{R}, \sigma} c_{j, \mathbf{R}+\rho, \sigma}. \qquad (14)$$

For simpler determination of the Hamiltonian matrix elements, we introduce vectors that follow the orientation of the hybrid orbitals: $\rho_1 = a_2$, $\rho_2 = a_1 - a_2$, and $\rho_3 = -a_1$. $\mathbf{R}$ and $a_i$ are the Bravais lattice translation vectors and primitive vectors, respectively. We also add the null vector $\rho_0 = 0$ useful in determining the on-site Hamiltonian matrix elements. The matrix elements that describe the electronic hopping between the nearest neighbors are shown in Fig. 1c. The diagonal (distant $\sigma$ bond) matrix elements are

$$H^{11}(\rho_1) = H^{22}(\rho_2) = H^{33}(\rho_3) = t_1. \qquad (15)$$

The diagonal (displaced $\pi$ bound) matrix elements are

$$H^{11}(\rho_3) = H^{22}(\rho_3) = H^{33}(\rho_2) = t_2. \qquad (16)$$

The off-diagonal matrix elements that represent the strongest hopping (mixture between the $\sigma$ and the $\pi$ bond) are

$$H^{12}(\rho_1) = H^{13}(\rho_1) = H^{32}(\rho_3) = t_3. \qquad (17)$$

The on-site matrix elements can be combined into the single expression

$$H^{ij}(0) = E_\varphi \delta_{ij} + (1 - \delta_{ij}) w_\phi. \qquad (18)$$

The diagonalization of Hamiltonian (14) provides three sigma bands $E_\mathbf{K}^{\sigma_i}$; $i = 1, 2, 3$ shown by blue dots in Fig. 2a, c. The TBA parameters used in the calculation are $t_1 = 0.39$, $t_2 = 0.2$, $t_3 = -0.55$, $E_\varphi = 1.9$, and $w_\varphi = -0.64$.

The TBA-ELF is calculated using the same procedure as described in the "Ab-initio approach" section except that, in (8), only $\sigma_1$ band is taken into consideration and all charge vertices (9) are approximated with unity ($\rho_{\sigma_1 \mathbf{K}} = 1$). This approach, called $\sigma_1$-TBA model, provides satisfactory good (in comparison with ab initio method) ELF in long-wavelength limit.

### DATA AVAILABILITY

The data generated during this study are available from the corresponding author on reasonable request.

### ACKNOWLEDGEMENTS

This work was supported by the QuantiXLie Centre of Excellence, a project cofinanced by the Croatian Government and European Union through the European Regional Development Fund—the Competitiveness and Cohesion Operational Programme (Grant KK.01.1.1.01.0004). I.L. was supported by the H2020 CSA Twinning Project No. 692194, RBI-T-WINNING. Computational resources were provided by the Donostia International Physics Center (DIPC) computing center.

### AUTHOR CONTRIBUTIONS

I.L. and V.D. devised the study and performed all DFT-based calculations. Z.R. performed tight binding calculations. All authors discussed the results and contributed to the final manuscript.

### ADDITIONAL INFORMATION

**Competing interests:** The authors declare no competing interests.

**Publisher's note:** Springer Nature remains neutral with regard to jurisdictional claims in published maps and institutional affiliations.